\documentclass[runningheads, envcountsame, a4paper]{llncs}
\usepackage{placeins}

\usepackage
	{todonotes}

\begin{document}

\title{Graph-based Features for Automatic Online Abuse Detection}
\titlerunning{Graph-based Features for Automatic Online Abuse Detection}

\author{Etienne Papegnies$^{1,2}$ \and Vincent Labatut$^1$ \and Richard Dufour$^1$ \and Georges Linar\`es$^1$}
\authorrunning{E. Papegnies, V. Labatut, R. Dufour, G. Linar\`es}

\institute{$^1$LIA -- EA 4128, University of Avignon, France \\ $^2$Nectar de Code, Barbentane, France \\ \email{firstname.lastname@univ-avignon.fr}}

\maketitle

\begin{abstract}
While online communities have become increasingly important over the years, the moderation of user-generated content is still performed mostly manually. Automating this task is an important step in reducing the financial cost associated with moderation, but the majority of automated approaches strictly based on message content are highly vulnerable to intentional obfuscation. In this paper, we discuss methods for extracting conversational networks based on raw multi-participant chat logs, and we study the contribution of graph features to a classification system that aims to determine if a given message is abusive. The conversational graph-based system yields unexpectedly high performance, with results comparable to those previously obtained with a content-based approach.
\keywords{Text categorization, Abuse detection, Online communities, moderation}
\end{abstract}

\section{Introduction}
\label{sec:Introduction}
The widespread availability of Internet access allows users from around the world to congregate into online communities. With the ever-increasing number of users, online communities have become important places to trade ideas, and have acquired a great socio-economical importance.

However, because of the anonymity provided by the medium, an online community is often confronted with users that display abusive behaviors. For community maintainers, it can be important to act on this issue through the use of moderation, because failure to do so can poison the community, trigger user exodus, and expose the administrators to legal jeopardy. Moderation is the application of sanctions when users are judged to violate the community rules. When done by humans, this work is expensive and companies have a vested interest in automating the process. One can distinguish two types of automated systems assisting in moderation: 1) an automated flagging system that raises some messages to the attention of moderators; and 2) a fully automated system that detects abusive messages and executes sanctions on users that are breaking the community rules.

In this work, we consider the classification problem of automatically determining if a message from a user is abusive or not. For this purpose, we propose an original approach aiming at exploring a range of graph-based features extracted from online textual conversations. We first extract various types of \textit{conversational networks}, {\it i.e.} graphs where vertices represent users and where edges correspond to supposed message-based interactions between them. We then process a number of graph-theoretical measures that characterize these networks in different ways. A classifier is then trained  and tested on a corpus of chat logs originating from the community of the French massively multiplayer on-line game \textit{SpaceOrigin}\footnote{\url{https://play.spaceorigin.fr/}}. We finally conduct a qualitative study to analyze the impact of each graph-based feature on the automatic abusive message classification performance. 

The rest of this paper is organized as follows. In Section~\ref{sec:Relwork}, we review related work on abuse detection and network extraction from raw conversation logs. In Section~\ref{sec:methods}, we describe the method proposed to extract conversational networks, and the topological features that we compute for the resulting graphs. In Section~\ref{sec:experiment}, our dataset is presented as well as the overall experimental setup for the classification task. A discussion and a qualitative study of our results is also provided. Finally, we summarize our contributions in Section~\ref{sec:Conclusion} and present some perspectives.

\section{Related Work}
\label{sec:Relwork}
This section is a brief review of the literature focusing on the most relevant works relating to two aspects of the problem at hand. First, in Subsection \ref{sec:rel-abuse}, we review general works regarding the detection of online abuse. Second, in Subsection \ref{sec:rel-network}, we explore previously used techniques to extract network structures from raw conversation data.

\subsection{Abuse Detection}
\label{sec:rel-abuse}
One can distinguish two main categories of works related to abuse detection: those using the content of the exchanged messages and those focusing on their context. Some works also propose to combine both categories. 

From the content-based point-of-view, the work initiated by Spertus in~\cite{spertus1997smokey} was a first attempt to create a classifier for hostile messages. This is relevant to us, because abusive messages often contain hostility. They use static rules to extract linguistic markers for each message: Imperative Statement, Profanity, Condescension, Insult, Politeness and Praise. These are then used as features in a binary classifier. They obtain good results, except in specific cases like hostility through sarcasm. However, the limitation of this approach is that its application to another language is a difficult task, requiring to transpose it to other grammar rules and idioms.

Cheng \textit{et al}. \cite{chen2012detecting} note that a word tagged as offensive in a message is not a definite indication that the message is offensive, i.e. while "You are stupid." is clearly offensive, "This is stupid. xD" is not. Lack of context can be somewhat mitigated by looking at word $n$-grams instead of unigrams (i.e. single words).

Dinakar \textit{et al}. \cite{dinakar2011modeling} use $tf$-$idf$ features, a static list of badwords and of widely used sentences containing verbal abuse to detect cyberbullying in Youtube comments. Again, their model showed good results, except when sarcasm was used. 

In \cite{chavan2015machine}, Chavan \textit{et al}. review machine learning approaches to detect aggressive messages in on-line social networks. They show that Pronoun Occurrence, usually neglected in text classification, is important, and use Skip-Gram features to mitigate the context issues.

Content-based text classification usually makes for a good baseline. Content features are inexpensive to compute. However, such methods have severe limitations: for instance, abuse can be spread over a succession of messages. Some messages can reference a shared history between two users. Even more common are users that are voluntarily obfuscating message content to work around badwords detection. Indeed, abusers can bypass automatic systems by making the abusive content difficult to detect \cite{hosseini2017deceiving}: for instance, they can intentionally modify the spelling of a forbidden word.

Because the reactions of other users to an abuse case are completely beyond the control of the abuser, some works consider the content of messages \textit{around} the targeted message. 

For instance, Yin \textit{et al}. \cite{yin2009detection} use features derived from the neighboring phrases of a given message to detect harassment on the Web. Their goal is to spot conversations going off-topic, and use that as an indicator. Their approach shows good results when used against multi-participant chat logs, and they note that sentiment features seem to constitute mostly noise due to the high misspelling rate.

In \cite{cheng2015antisocial}, Cheng \textit{et al}. propose to focus on building user behavior models.   For this purpose, they perform a comprehensive study of antisocial behavior in on-line discussion communities. Their work provides insight into the devolution of abusive users over time in a community, regarding both the quality of their contributions and their reactions towards other members of the community. A critical result of the analysis is that instances of antisocial messages usually generate a bigger response from the community, compared to normal messages.

Balci \textit{et al}. \cite{balci2015automatic} make use of user features to detect abuse in the community of an online game. These features include information such as gender, number of friends, financial investment, avatars, and general rankings. The goal is to help human moderators dealing with abuse reports, and the approach yields sufficiently good results to achieve it. However, in our case the user data necessary to replicate this approach is not available.

In our own previous work \cite{papegnies2017impact}, we propose to detect abusive messages from chat messages using a wide array of language features (bag-of-words, $tf$-$idf$ scores, sentiment scores, etc.) as well as context features derived from the language models of other users. We also try advanced preprocessing approaches. This method allowed us to reach a performance of 72.1\% in terms of $F$-measure on an abusive message detection task.

\subsection{Network Extraction from Raw Data}
\label{sec:rel-network}
Although a major part of the solutions focus on content features of exchanged messages to address the abuse problem, it appears that a user with previous exposure to automatic moderation techniques can easily circumvent them~\cite{hosseini2017deceiving}. To avoid this problem, a solution would be to not focus only on the direct content exchanged but on the interactions between the users through these messages.

The number of respondents to a given message appears frequently in the literature, as a classification feature, e.g. \cite{cheng2015antisocial}. However, there are not many works dealing with the extraction of conversational networks. This may be due to the fact that the task can be far from trivial, depending on the nature of the available raw data: the task is much harder for chat logs than for structured messages board or Web forums, for instance. These networks have the advantage of including the mentioned feature, but also much more information regarding the way users interact.

In \cite{mutton2004inferring}, Mutton proposes a strategy to extract such a network from IRC chat logs. The goal is to build a tool to visualize user interactions in an IRC chat room over time. The author uses a simple set of rules based on \textit{direct referencing} (when a user addresses another one by using his nickname), temporal proximity of messages, and temporal density of messages. In this paper, we will adapt and expend on those rules. Specifically, while in a regular IRC channel timestamps are indeed useful to determine intended recipients of a message, in our case they are basically irrelevant, so this approach cannot be adapted as is.

Travassoli \textit{et al}. \cite{tavassoli2014constructing} explore different methods to extract representative networks from group psychotherapy chat logs. One method includes fuzzy referencing to mitigate effects of misspelled nicknames, and rules for representing one-to-all messages. The bulk of the methods uses static patterns of exchanges to predict a receiver. Their system shows a good agreement score with a human annotator.

Sinha \textit{et al}. \cite{sinha2014investigating} use only direct referencing, but with the same fuzzy matching strategy, in order to extract a network representing the activity in the \texttt{\#ubuntu} IRC support channel. This method manages to expose high level components of the Ubuntu social network, which in turn allows for the qualification of user behaviors into specific classes. This method of building user models can be very interesting when the data describing the users are scarce, as is the case on IRC where everyone can join and there is no requirement to register.

\section{Methods}
\label{sec:methods}
In this section, we describe our proposed original approach to detect abusive messages. It basically consists in training a classifier on features corresponding to topological measures processed on conversational networks. The classifier is standard, so we focus on the processing of the features. Thereby, Subsection~\ref{sec:experiment-networkinference} presents how we extract conversational networks from conversation logs, while Subsection~\ref{s:features} describes the topological measures computed for these conversational networks, and later used as classification features.

\subsection{Network Extraction}
\label{sec:experiment-networkinference}
We extract networks representing conversations between users, through a textual discussion channel. They take the form of weighted undirected graphs, in which the vertices and edges represent the users and the communication between them, respectively. The edge weights are a score which is an estimation of the intensity of the communication between the two connected users. Note that each network is defined relatively to a \textit{targeted message}, since the goal of this operation is to provide features used to classify the said message.

The first step consists in determining which messages are used to extract the network. For this purpose, we define a \textit{context period}, which is centered on the \textit{targeted message}, and spans symmetrically before and after its occurrence. We arbitrarily use a width of $200$ messages in our experiments. The graph extracted from this context period contains only the vertices representing the users which posted at least once on this channel, during this period.

The second step is to add the appropriate edges to the network, and to process their weight. We use a method based on a sliding window, a choice that is justified by two properties of the user interface of the considered discussion channel: 1) when a user joins a channel, the server sends him only the last $20$ messages posted on the channel; and 2) it is impossible for a user to scroll back the history further than $20$ lines. In our experiments, we arbitrarily use a window of $10$ messages. We apply an iterative process, consisting in sliding the window over the whole context one message at a time. We call \textit{current message} the \textit{last} message of the window taken at a given time. Our assumption is that this message is destined to the authors of the other messages present in the window at this time. Furthermore, we suppose it is more likely that the message concerns the users who posted the most recently. These hypotheses can be justified by another property of the user interface: by default, users do not know who is in the channel at a given time, in particular the join / part events are not shown to them. 

Based on these hypotheses, we update the edges and weights in the following way. First, we list the authors of the messages currently present in the window, order them by last message posted, and discard the author of the \textit{current message} (since it is possible that several of his messages appear in this window): this results in what we call the \textit{neighbor list}. However, the user interface allows to \textit{explicitly} mention users in a message by their name, and the game prevents the users from changing their name: we need to take this property into account. For this purpose, we move the users directly referenced in the \textit{current message} at the top of our list. If a user was not even in the window, it is simply inserted at the top of the list. Each user in the neighbor list is assigned by a score, which is a decreasing function of both his position in the list and of the length of the \textit{neighbor list}. We can then update the graph: we create an edge between each user in the neighbor list and the author of the \textit{current message}, with a weight corresponding to the user's score. If this edge already exists, we increase its current weight by the user's score.

Our choice to create or update edges towards all users in the window even in case of direct referencing is based on several considerations. First, directly referencing a user does not imply that he is part of the conversation or that the message is directed towards him: for instance, his name could just be mentioned as an object of the sentence. Second, there can be multiple direct references in a single message. Third, when in online public discourse, directly addressing someone does not mean he is the sole intended recipient of the message. For instance when discussing politics, a question directed towards someone can have as a secondary objective to have the target expose his stance on an issue to the other participants.

\begin{figure}[!ht]
    \center
    \resizebox{\textwidth}{!}{
	    \includegraphics[width=0.333\textwidth]{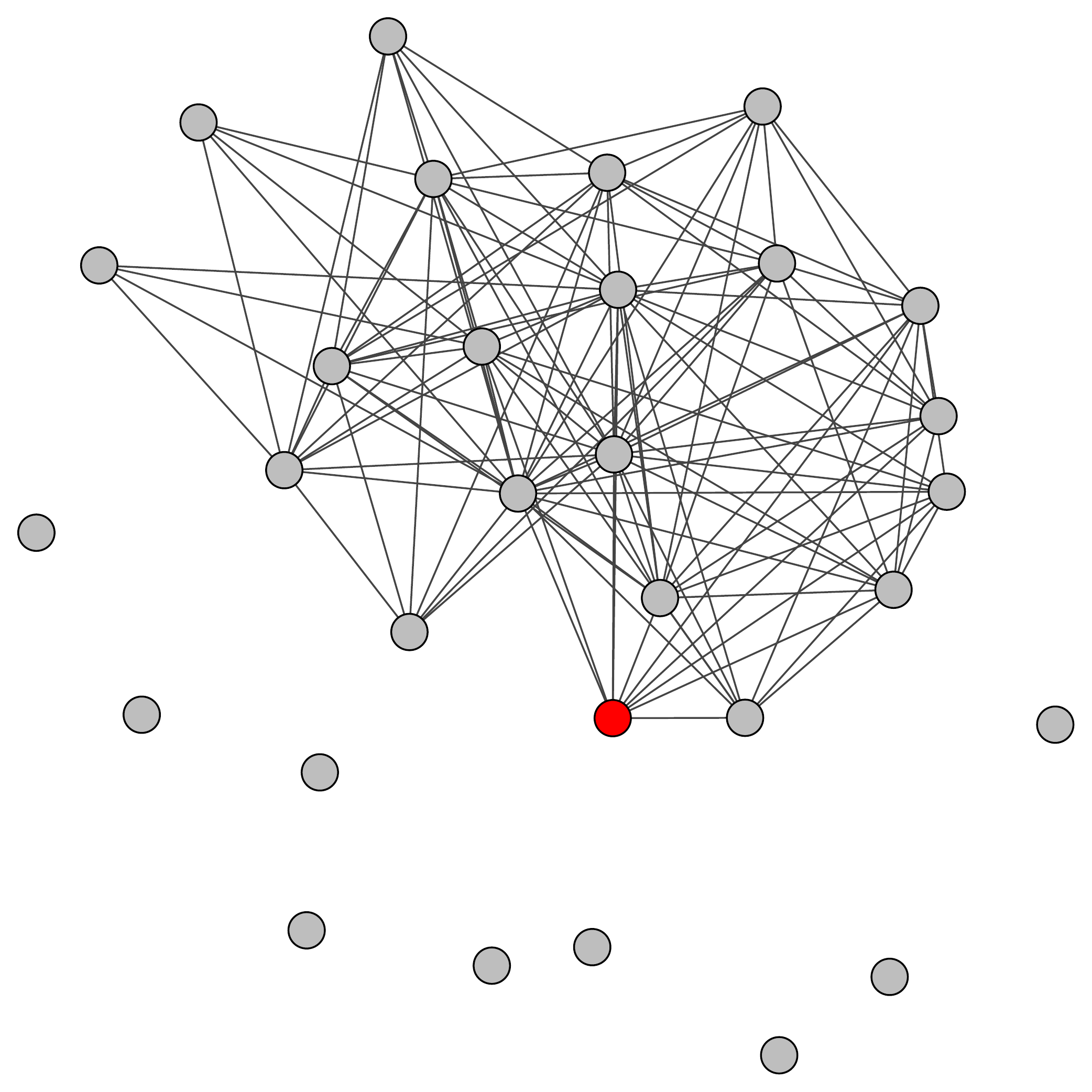}\hfill
	    \includegraphics[width=0.333\textwidth]{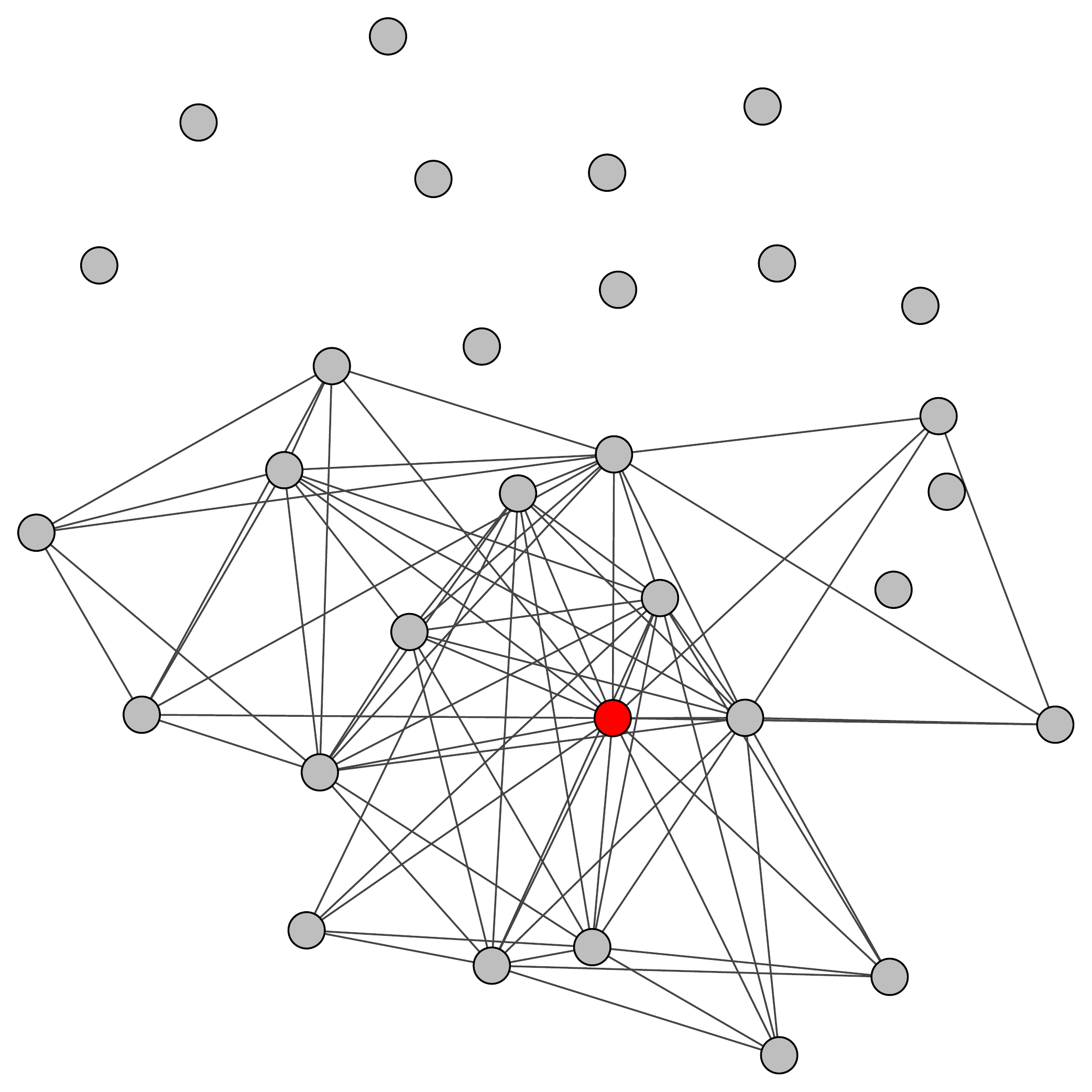}\hfill
	    \includegraphics[width=0.333\textwidth]{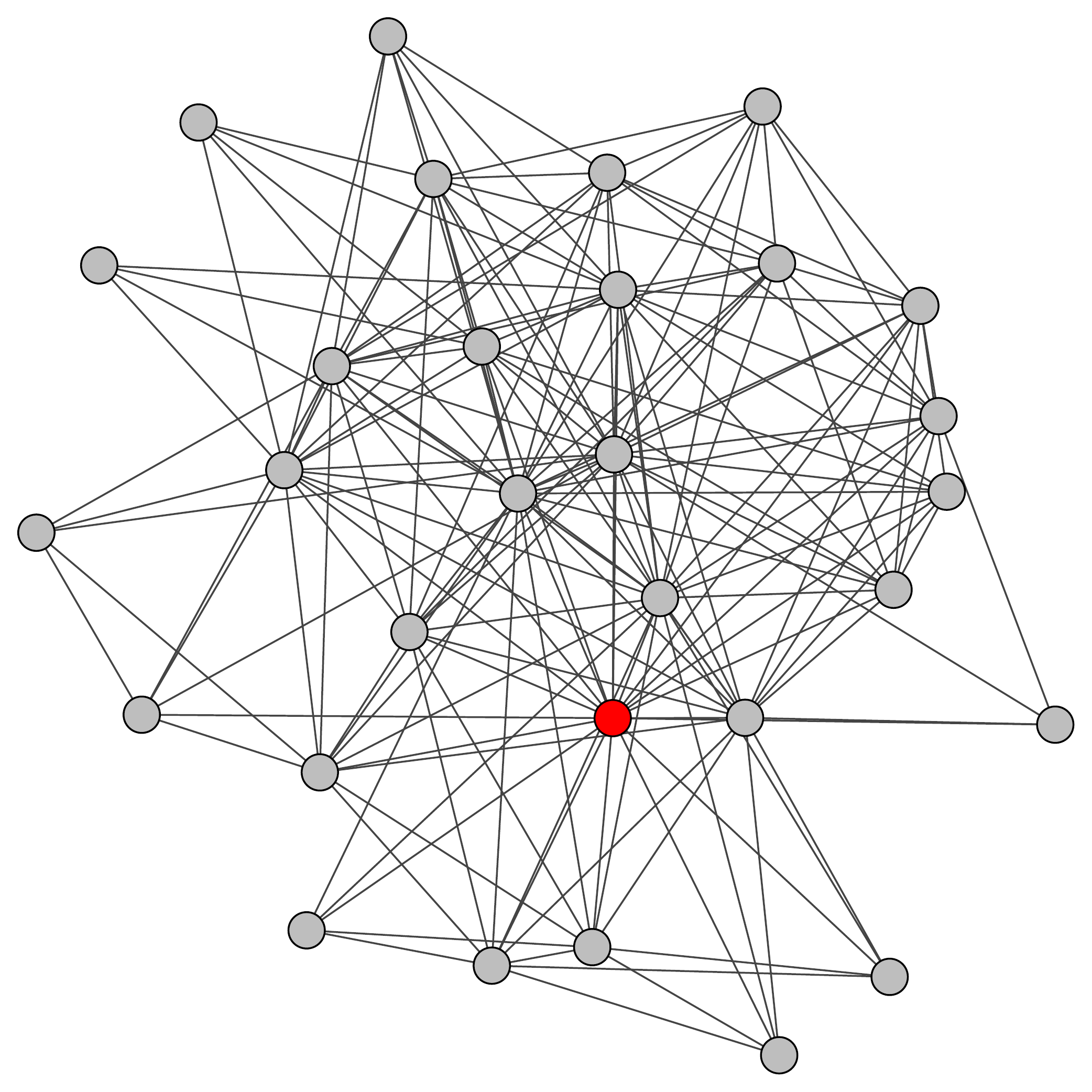}
    }
    \caption{Example of the $3$ types of conversational networks extracted for a given context period: Before (left), After (center), and Full (right). The abusive user is represented in red.}
	\label{fig:nets}
\end{figure}

Once the iterative process has been applied for the whole context period, we get what we call the \textit{Full} network. For testing matters, we also process $2$ lesser networks based on the same context: the \textit{Before} and \textit{After} networks are extracted using only the $100$ messages preceding and following the \textit{targeted message}, respectively, as well as the \textit{targeted message} itself. Figure~\ref{fig:nets} shows an example of the three networks associated with an abusive comment.

\FloatBarrier

\subsection{Features}
\label{s:features}
The classification features we consider in this work are all topological measures, allowing to characterize graphs in various ways. We adopt an exploratory approach and consider a wide range of such measures, focusing on the most widespread in the literature. In the following, we describe them briefly, distinguishing between \textit{local} ones, which characterize individual vertices, and \textit{global} ones, which describe the whole graph at once. We process all the features for each of the $3$ types of networks (Before, After, Full) described in the previous subsection.

\subsubsection{Local Topological Measures.} These measures are computed for the Vertex corresponding to the author of the targeted message. 

The \textit{Degree centrality} is a normalized version of the standard degree, which corresponds itself to the number of direct neighbors of the considered vertex. The \textit{Eigenvector Centrality} \cite{Bonacich1987} can be considered as a generalization of the degree, in which instead of just counting the neighbors, one also takes into account their centrality: a central neighbor increases the centrality of the vertex of interest more than a peripheral one. 

The \textit{PageRank Centrality} \cite{Brin1998} is also spectral (like the  \textit{Eigenvector Centrality}), but it is based on very different rationales. It models a random walk occurring on the network, and noticeably includes the possibility for the walker to teleport anywhere in the network at any step. The \textit{Hub} and \textit{Authority Scores} \cite{Kleinberg1999} are two complementary measures also based on random walks. 

The \textit{Betweenness Centrality} \cite{Freeman1978} is based on the number of shortest paths going through the considered vertex. 

In communication networks, it is sometimes interpreted as the level of control the vertex of interest has over information transmission in the network. The \textit{Closeness Centrality} \cite{Freeman1978} is the reciprocal of the total geodesic distance (i.e. the length of the shortest path) between the vertex of interest and the other vertices. It is generally considered it measures the efficiency of the vertex to spread a message over the graph, and its independence from the other vertices in terms of communication. The \textit{Eccentricity} \cite{Harary1969} is also distance-based, but to the contrary of the other selected measures, it quantifies how peripheral the vertex of interest is, by considering the distance to its farthest vertex.

Finally, the \textit{Coreness Score} \cite{Seidman1983} is based on the notion of $k$-core, which is a maximal induced subgraph whose all vertices have a degree of at least $k$. The coreness of a vertex is the $k$ of the $k$-core of maximal degree to which it belongs.

\subsubsection{Global Topological Measures.}
First, we use very classic statistics describing the graph size, the \textit{Vertex} and \textit{Edge Counts}. We also select the \textit{Density}, which corresponds to the ratio of the number of existing edges to the number of edges in a complete graph containing the same number of vertices. In other words, the density corresponds to the proportion of existing edges, compared to the maximal possible number for the considered graph.

We also use two distance-related measures. The first is the \textit{Diameter}, which corresponds to the highest distance found in the graph, i.e. the length of the longest shortest path. The second is the \textit{Average Distance}, which is the average length of the shortest path processed over all pairs of vertices.

We process the total \textit{Clique Count} in the network, where a clique is a complete induced subgraph. The \textit{Degree Assortativity} \cite{Newman2002} is also potentially interesting. It corresponds to the correlation processed between the series constituted of all connected vertices, and measures the statistical dependence between the degrees of two vertices and the presence of an edge connecting them. Finally, for each one of the $10$ previously described local measures, we process the average over the whole graph.

\section{Experiments}
\label{sec:experiment}
In this section, we first briefly present our corpus and our experimental setup (Subsection~\ref{sec:experiment-setup}), before describing and discussing our classification results (Subsection~\ref{sec:experiment-results}).


\subsection{Experimental Setup}
\label{sec:experiment-setup}
We have access to a database of $4,029,343$ messages that were exchanged by the users of a browser-based multi-player game. In the database, $779$ messages have been flagged by one or more users as being abusive and subsequently confirmed as abusive by the game moderators. Each message belongs to a unique communication channel.

We further extract $2,000$ messages at random from the messages not confirmed as abusive to constitute the non-abuse class. We previously experimented with this dataset in \cite{papegnies2017impact}.


Because of the relatively small dataset, our experiment is set up for $10$-Fold cross validation.
We use a 70\%-train / 30\%-test split.

We use Python-iGraph \cite{csardi2006igraph} to create the network and process the graph-based features for each message. As a classifier, we use an SVM, implemented in Sklearn under the name SVC (C-Support Vector Classification) \cite{pedregosa2011scikit} toolkit.

\subsection{Results}
\label{sec:experiment-results}
Table \ref{fig:res} displays the results obtained for our random baseline, the content- and context-based classifier we previously presented in \cite{papegnies2017impact}, and the graph-based classifier proposed in this article. The baseline uses the same classifier and architecture but the feature extraction step is replaced by a dummy function that yields two random values in $[0, 1]$. Our previous approach takes advantage of morphological, language and user behavior-based features, such as: message length, number of words, compressibility, bag of words with $tf$--$idf$ scores and probability of $n$-gram emission. In the present experiment, the training and testing sets were resampled, which explains why the values displayed for the content/context-based classifier are slightly different form the ones shown in \cite{papegnies2017impact}.

With the graph-based approach, the performance is improved according to all $3$ considered measures, compared to our previous effort. The overall performance is unexpectedly high for an approach that \textit{completely ignores the content} of the messages. We suppose that this is mainly due to the fact that two thirds of the features include information regarding to the part of the conversations happening \textit{after} the classified message, whereas this was the case for only \textit{two} features (out of 67) in our content/context-based approach. Independently from this point, the fact that both approaches reach relatively high performance levels is a very promising result: given that both classifiers are built on completely different features, combining them should even improve the overall performance.

\begin{table}[htp]
	\center
	\caption{Classification results (in \%) of the $3$ abusive message classifiers: a random baseline, our previous approach \cite{papegnies2017impact}, and the one presented in this article. All measures are computed for the \textit{Abuse} class.}
	\begin{tabular}{ p{6.3cm}@{\hskip 0.5cm}r@{\hskip 0.5cm}r@{\hskip 0.5cm}r }
		\hline
    	\textbf{Experiment} & \textbf{Precision} & \textbf{Recall} & \textbf{$F$-Measure} \\ 
        \hline 
        \textbf{Random baseline} & 29.3 & 52.6 & 37.6 \\
        \textbf{Content/context-based classifier} & 70.3 & 74.3 & 72.2 \\
       	\textbf{Graph-based classifier} & 76.8 & 77.2 & 77.0 \\
        \hline 
	\end{tabular}
	\label{fig:res}
\end{table} 

\begin{figure}[htp]
    \resizebox{\textwidth}{!}{
	    \includegraphics[width=0.499\textwidth]{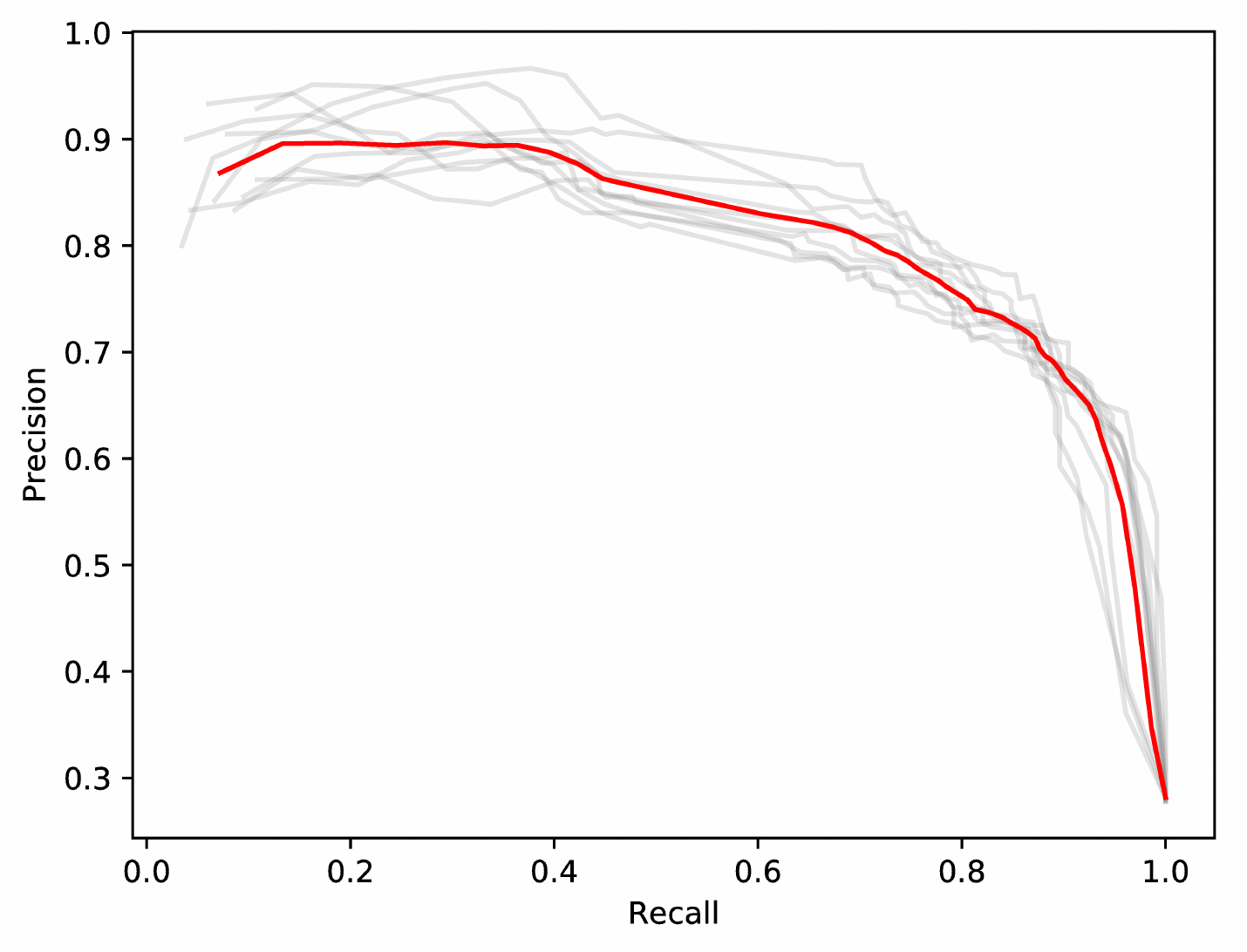}
        \includegraphics[width=0.499\textwidth]{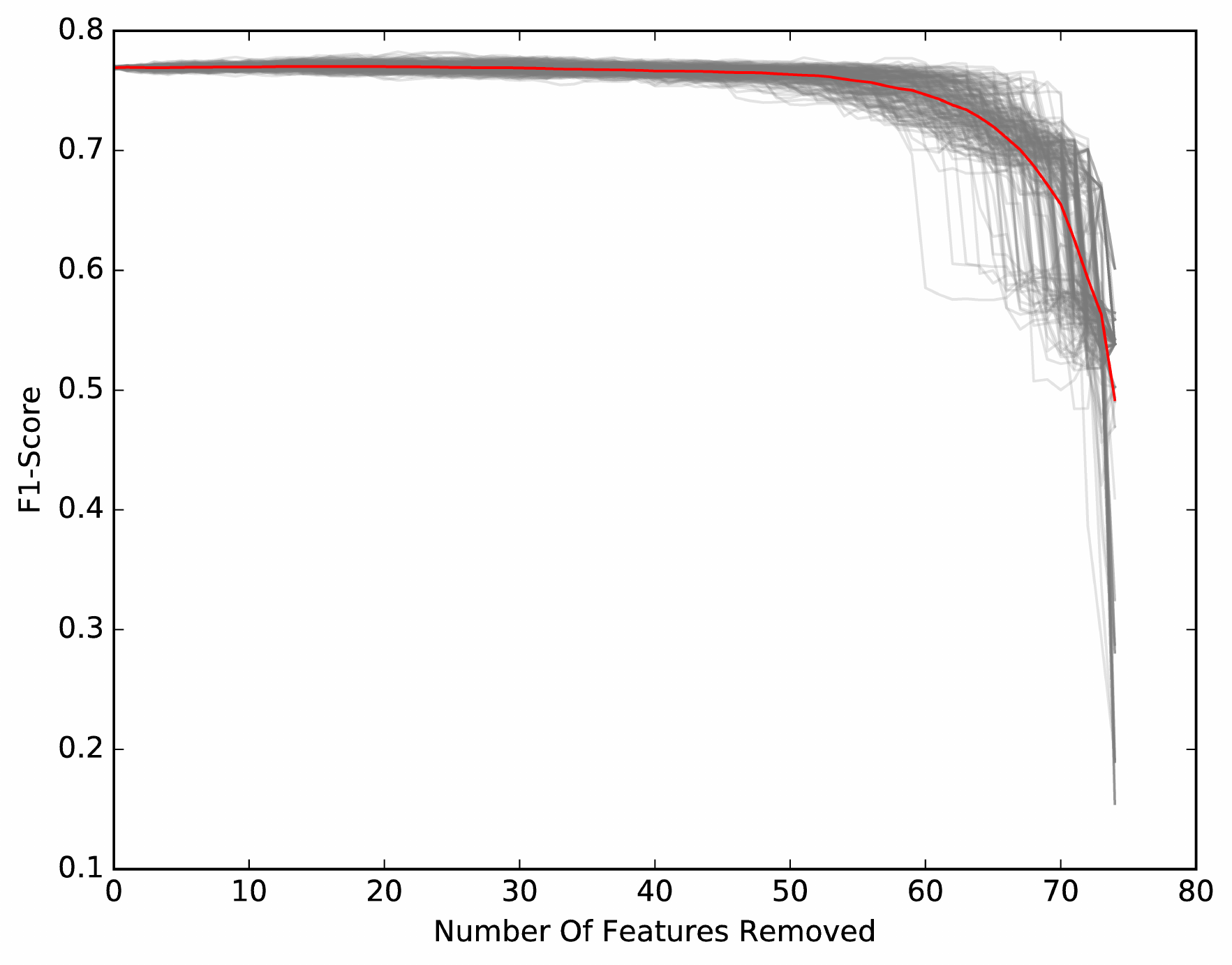}
    }
	\caption{Left: Precision-Recall curves of the $10$ SVM classifiers. Right: Feature ablation curves of one classifier (200 runs). In both plots, the red curve represents the average.}
	\label{fig:Curve}
\end{figure}

Since our classifier is an SVM, we can use the Platt Scalling implementation of Sklearn to vary the decision threshold and therefore tune the system towards either high precision or high recall. The left plot of Figure~\ref{fig:Curve} shows the Precision-Recall curves of each of the $10$ classifiers created for our experiment. One can see that by lowering the post probability threshold a little, it is possible to gain better coverage of the abuse class without losing too much precision. Therefore, we would argue that this system shows better promise as an alert system than as an automated moderation system.

In order to estimate the importance of our features with regards to this classification task, we use the meta estimator \textit{ExtraTreesClassifier} provided by the Sklearn toolkit. While the process is stochastic, it allows to give features a score indicating their contribution to the decisions of the classifier. We run further ablation runs, ordering the features by increasing impact of their removal on the classifier performance. This allows obtaining a smoother curve, with a performance drop on the right-side, corresponding to the removal of the most discriminant features (right plot of Figure~\ref{fig:Curve}). Table \ref{tableTop} shows the $10$ most discriminant features: using only these features, one can train a classifier obtaining a $F$-Measure score of $75.8\%$.
\begin{table}[htp]
	\center
    \caption{Most useful features of the graph-based classification approach.}
    \begin{tabular}{ l@{\hskip 0.5cm}p{4.8cm}@{\hskip 0.5cm}r@{\hskip 0.5cm}r }
		\hline
     	\textbf{Network} & \textbf{Feature name} & \textbf{$F$-Measure \textit{before} ablation} \\
        \hline
        Full & Average Betweenness Centrality & 75.8 \\
        Before & Average Coreness Score & 75.4 \\
        After & Edge Count & 74.5 \\
        After & Density & 73.1 \\
        Full & Hub Score & 72.9 \\
        After & Degree Centrality & 67.7 \\
        Before & Edge Count & 67.2 \\
        Full & Average Eccentricity & 58.4 \\
        Before & Average Eigenvector Centrality & 56.6 \\
        Full & Eccentricity & 35.0 \\
        \hline
  	\end{tabular}
	\label{tableTop}
\end{table} 

Overall, these features are quite heterogeneous, topologically speaking, in the sense they correspond to very different ways of characterizing graph structures. The \textit{Degree Centrality}, \textit{Edge Count}, and \textit{Density} features are based on a microscopic view of the graph (vertices and edges are considered individually, or only with respect to their direct neighborhood). On the contrary, the \textit{Betweenness Centrality}, \textit{Hub Score} and \textit{Eccentricity} are macroscopic, because they take advantage of paths spanning the whole graph. Finally, the \textit{Coreness Score} is mesoscopic, in the sense it is based on an intermediate view and considers subgraphs. This is consistent with the assumption that redundant features should not appear amongst the most discriminant ones. 

At first sight, finding both \textit{Edge Count} and \textit{Density} can be surprising: given that the latter is a normalized version of the former, one could suppose they are redundant. However, this normalization is based on the number of vertices in the graph. Thus, in the present case, this simply means that the number of edges in our networks does not increase as a square function of number of vertices. On the contrary, certain features present in the table are part of some very correlated groups of features, which can be considered as almost inter-exchangeable. For instance, the \textit{Average Eigenvector Centrality} and \textit{Average Hub Score} for the \textit{Before} graph have a $0.73$ correlation.

All $3$ considered types of graphs (\textit{Before}, \textit{After}, \textit{Full}) are represented in these top features, which means they convey different information and are all of some help regarding the classification problem at hand. Moreover, it appears that certain related features appear together for several versions of the graph. This is the case for the \textit{Edge Count} (\textit{Before} vs. \textit{After}), and of the Hub Score and Eigenvector Centrality (\textit{Full} vs. \textit{Before}). We assume that this reflects the fact abuses significantly modify the graph structure, according to these topological measures. In other words, they reflect strong changes in the conversation dynamics. When a measure appears only for the \textit{Before} or \textit{After} version of the graph, we conclude it allows characterizing only the pre- or post-state of the conversation, relatively to the abuse.

It is interesting that both the individual and average Eccentricity features are present in this table. A closer look reveals that their values are lower for graphs belonging to the \textit{Abuse} class. This means that the maximal distance between the author of the targeted message and the rest of the graph decreases in case of abuse. More concretely, this user becomes less peripheral (or more central), and the same goes for the other users of the graph (in average). This fits in quite well with assumptions about how abuse impacts a discussion: an abuser would tend not to be peripheral in a conversation, while we can reasonably assume that the other participants will be piling on and therefore be less peripheral themselves. This may also explain why those features are, by far, the most discriminant ones.

\section{Conclusion}
\label{sec:Conclusion}
In this article, we have presented an approach purely based on graph features to tackle the problem of automatically detecting online abuse. The method, while simple, yields reasonable results, besting the score obtained with our previous effort, which was content- and context-based. 

However it is important to note a couple of important limitations. First, the amount of necessary computation is quite high if it is to be applied each time a new message is posted to the channel, compared to a pure content-based approach. Second, the method can only be applied after a delay when the necessary number of messages have been posted in response to the target message - this is not a method that can help \textit{prevent} the $5^{th}$ message in a torrent of insults from reaching the channel. Rather, it could be used to perform some \textit{a posteriori} moderation.

The next step in our study will be to assess the impact of different network construction strategies on the performance of the classifier. This will include experimenting with other weight distribution strategies, and different sizes for the context period and the sliding window. We will then aim to combine this system with our content-based classifier: in theory, they are both based on completely different types of information, so we can assume they are complementary and could lead to improved classification performance.

\section*{Acknowledgments.}
This work was financed by a grant from the \textit{Provence Alpes Côte d'Azur} region (France) and the \textit{Nectar de Code} company.

\bibliographystyle{splncs03}
\bibliography{TBD}

\begin{thebibliography}{10}
\providecommand{\url}[1]{\texttt{#1}}
\providecommand{\urlprefix}{URL }

\bibitem{balci2015automatic}
Balci, K., Salah, A.A.: Automatic analysis and identification of verbal
  aggression and abusive behaviors for online social games. Computers in Human
  Behavior  53,  517--526 (2015)

\bibitem{Bonacich1987}
Bonacich, P.F.: Power and centrality: A family of measures. American Journal of
  Sociology  92,  1170--1182 (1987)

\bibitem{Brin1998}
Brin, S., Page, L.E.: The anatomy of a large-scale hypertextual web search
  engine. Computer Networks and ISDN Systems  30,  107--117 (1998)

\bibitem{chavan2015machine}
Chavan, V.S., Shylaja, S.S.: Machine learning approach for detection of
  cyber-aggressive comments by peers on social media network. In: IEEE ICACCI.
  pp. 2354--2358 (2015)

\bibitem{chen2012detecting}
Chen, Y., Zhou, Y., Zhu, S., Xu, H.: Detecting offensive language in social
  media to protect adolescent online safety. In: PASSAT/SocialCom. pp. 71--80
  (2012)

\bibitem{cheng2015antisocial}
Cheng, J., Danescu-Niculescu-Mizil, C., Leskovec, J.: Antisocial behavior in
  online discussion communities. preprint arXiv:1504.00680  (2015)

\bibitem{csardi2006igraph}
Csardi, G., Nepusz, T.: The igraph software package for complex network
  research. InterJournal Complex Systems  1695(5),  1--9 (2006)

\bibitem{dinakar2011modeling}
Dinakar, K., Reichart, R., Lieberman, H.: Modeling the detection of textual
  cyberbullying. The Social Mobile Web  11, ~02 (2011)

\bibitem{Freeman1978}
Freeman, L.C.: Centrality in social networks i: Conceptual clarification.
  Social Networks  1(3),  215--239 (1978)

\bibitem{Harary1969}
Harary, F.: Graph Theory. Addison-Wesley (1969)

\bibitem{hosseini2017deceiving}
Hosseini, H., Kannan, S., Zhang, B., Poovendran, R.: Deceiving google's
  perspective api built for detecting toxic comments. preprint arXiv:1702.08138
   (2017)

\bibitem{Kleinberg1999}
Kleinberg, J.: Authoritative sources in a hyperlinked environment. Journal of
  the Association for Computing Machinery  46(5),  604--632 (1999)

\bibitem{mutton2004inferring}
Mutton, P.: Inferring and visualizing social networks on internet relay chat.
  In: 8th International Conference on Information Visualisation. pp. 35--43
  (2004)

\bibitem{Newman2002}
Newman, M.E.J.: Assortative mixing in networks. Physical Review Letters
  89(20),  208701 (2002)

\bibitem{papegnies2017impact}
Papegnies, E., Labatut, V., Dufour, R., Linares, G.: Impact of content features
  for automatic online abuse detection. In: International Conference on
  Computational Linguistics and Intelligent Text Processing (2017)

\bibitem{pedregosa2011scikit}
Pedregosa, F., Varoquaux, G., Gramfort, A., Michel, V., Thirion, B., Grisel,
  O., Blondel, M., Prettenhofer, P., Weiss, R., Dubourg, V., et~al.:
  Scikit-learn: Machine learning in python. Journal of Machine Learning
  Research  12,  2825--2830 (2011)

\bibitem{Seidman1983}
Seidman, S.B.: Network structure and minimum degree. Social Networks  5(3),
  269--287 (1983)

\bibitem{sinha2014investigating}
Sinha, T., Rajasingh, I.: Investigating substructures in goal oriented online
  communities: Case study of ubuntu irc. In: IEEE International Advance
  Computing Conference. pp. 916--922 (2014)

\bibitem{spertus1997smokey}
Spertus, E.: Smokey: Automatic recognition of hostile messages. In: 14th
  National Conference on Artificial Intelligence and 9th Conference on
  Innovative Applications of Artificial Intelligence. pp. 1058--1065 (1997)

\bibitem{tavassoli2014constructing}
Tavassoli, S., Moessner, M., Zweig, K.A.: Constructing social networks from
  semi-structured chat-log data. In: IEEE/ACM International Conference on
  Advances in Social Networks Analysis and Mining. pp. 146--149 (2014)

\bibitem{yin2009detection}
Yin, D., Xue, Z., Hong, L., Davison, B.D., Kontostathis, A., Edwards, L.:
  Detection of harassment on web 2.0. In: WWW Workshop: Content Analysis in the
  WEB 2.0 (2009)

\end{thebibliography}

\end{document}